\documentclass[letterpaper, 10 pt, conference]{ieeeconf}  % Comment this line out
\IEEEoverridecommandlockouts                              % This command is only
\usepackage{amsmath} % assumes amsmath package installed
\usepackage{graphicx}
\usepackage{mathrsfs}
\usepackage{color}
\usepackage{hyperref}
\usepackage{ctable}
\usepackage{multirow}

\newcommand{\otoprule}{\midrule[\heavyrulewidth]}
\title{\LARGE \bf
Multi-Community Detection in Signed Graphs \\ Using Quantum Hardware
}

\author{ \parbox{6.8 in}{\centering Ehsan Zahedinejad,$^1$ Daniel Crawford,$^1$ Clemens Adolphs,$^1$ and Jaspreet S. Oberoi$^{1,2,*}$\\
        $^1$1QB Information Technologies (1QBit),\\
        458-550 Burrard Street, Vancouver, BC, V6C 2B5, Canada\\
        $^2$School of Engineering Science, Simon Fraser University, \\ 8888 University Drive, Burnaby, BC, V5A 1S6, Canada\\ 
        {\tt\small $^*$jaspreet.oberoi@1qbit.com}}
        \hspace*{ 0.025 in}
}

\begin{document}

\maketitle
\thispagestyle{plain}
\pagestyle{plain}

%%%%%%%%%%%%%%%%%%%%%%%%%%%%%%%%%%%%%%%%%%%%%%%%%%%%%%%%%%%%%%%%%%%%%%%%%%%%%%%%
\begin{abstract}
Signed graphs serve as a primary tool for modelling social networks. They can represent relationships between individuals (i.e., nodes) with the use of signed edges. Finding communities in a signed graph is of great importance in many areas, for example, targeted advertisement. We propose an algorithm to detect multiple communities in a signed graph. Our method reduces the multi-community detection problem to a quadratic binary unconstrained optimization problem and uses state-of-the-art quantum or classical optimizers to find an optimal assignment of each individual to a specific community.
\end{abstract}

%%%%%%%%%%%%%%%%%%%%%%%%%%%%%%%%%%%%%%%%%%%%%%%%%%%%%%%%%%%%%%%%%%%%%%%%%%%%%%%

\section{Introduction}
Signed graphs (SG) are ubiquitous in social networks~\cite{MA05,LHK10,KLB09}. They can encode the perception and attitude between individuals via signed links, where a positive link between two nodes can indicate friendship and trust, while a negative link denotes animosity and distrust~\cite{Fri46}. Thus far, there has been impressive progress towards the development of methods for exploring  tasks within SGs~\cite{BCM11,LK03,AS14}. With the continuous rapid yearly growth of social media users, there is an immediate need for reliable and effective approaches to the modelling of social networks.

There exists a range of interesting problems within the SG domain, including link prediction~\cite{LK03,CNT+11}, network evolution~\cite{AS14}, node classification~\cite{TAL16}, and community detection. In this work, we focus on developing a multi-community detection algorithm~\cite{TCA+16,FH93,SD94,Piz09,AM12}. The principal idea in community detection is to divide an SG into clusters such that the representations of users within a cluster are densely connected by positive links whereas those that belong to different clusters are connected by negative links. Community detection has numerous applications in various areas, including in medical science~\cite{CZG+12,SJ10}, telecommunications~\cite{FMC+14}, the detection of terrorist groups~\cite{Tod12}, and information diffusion processes~\cite{SQG+15}. The wide applicability of community detection makes it an important topic of study, and emphasizes the need to devise faster and more-effective approaches in its implementation.

Community detection research can be divided into four categories~\cite{TCA+16}, that is, clustering based, mixture-model based, dynamic-model based, and modularity based. Our approach is modularity based: we maximize modularity (i.e., the number of edges that fall within clusters minus the expected number of edges within those clusters in an equivalent network with edges placed at random) and minimize frustration~\cite{FH93} (i.e., the number of negative edges within communities plus the number of positive links between communities) to discover communities in an SG.

Over the last decade, there has been a large body of work that has used modularity or a variant of it as a metric for detecting communities in an SG. For instance, the authors of~\cite{AM12} find communities by minimizing the frustration and those of~\cite{AP13} propose a community detection framework called SN-MOGA, using a non-dominated sorting genetic algorithm~\cite{SD94,Piz09} to simultaneously minimize frustration and maximize signed modularity. Authors in~\cite{EAJ14} investigate the role of negative links in SGs that use frustration as a metric.

We formulate the multi-community detection problem as a quadratic unconstrained binary optimization (QUBO) problem, the optimal solution of which corresponds to the solution of the multi-community detection problem. Our approach has several advantages over  existing community detection algorithms. Unlike other approaches, our approach does not require an input from the user predefining the number of communities to be detected. It  requires only an upper-bound on the number of communities in order to find the optimal number of communities. Also, in the case of finding more than two communities, our approach does not recursively divide the graph into two parts~\cite{AM12}, creating artificial local boundaries between communities, but instead preserves the global structure of the SG. Lastly, our approach is applicable to any community detection metric which can be formulated as a QUBO problem.

In this work, we use two  solvers to address a QUBO problem, a classical algorithm called parallel tempering Monte Carlo with isoenergetic cluster moves (PTICM)~\cite{Mac98,HN96}, also known as ``borealis''~\cite{ZOK15, ZFK16}, and a quantum annealer (the D-Wave 2000Q~\cite{JAG+11}). We have chosen PTICM because of its proven superiority over other QUBO solvers~\cite{ZOK15,ZFK16}, while experiments with the D-Wave device have helped us to understand that our approach can immediately benefit from advancements in quantum computing~\cite{RHI+18}.

The current quantum annealer's processor has a small number of qubits, limiting the size of benchmark datasets we can use to test our algorithm. We have considered two approaches to remedy this constraint. First, trivially enough, we limit the maximum size of the selected dataset by choosing an SG with less than 64 nodes (this results in a QUBO problem with 256 binary variables). Second, we implement block coordinate descent (BCD) to enable our approach to find communities in larger SGs~\cite{ZHT15,RVW+16}. BCD works by iteratively solving subproblems while keeping the rest of the variables fixed. The performance of BCD on several QUBO problems is reported in~\cite{RVW+16}. 

The main focus of this work is to propose a new algorithmic approach for finding multiple communities in an SG. As such, we do not intend to compare and benchmark the performance of different QUBO solvers for solving this problem. The paper is structured as follows. We present  terminology and notation in Section~\ref{sec:notation}. We then explain the concept of structural balance its relation to community detection Section~\ref{sec:structural_balance}. We give our proposed method for multi-community detection in Section~\ref{sec:methods}. We explain our approach in comparing the performance of the proposed algorithms and our choice of benchmarking datasets in Section~\ref{sec:approach}. We report our results in Section~\ref{sec:results} and discuss them in Section~\ref{sec:discussion}. We conclude and suggest future research directions in Section~\ref{sec:conclusion}.

\vspace{0.5em}

\section{Notation}
\label{sec:notation}
In this section, we present the notation we use throughout this work. In our terminology, ${\textbf{G}}(\text{V},\text{E})$ denotes an SG where $\text{V}$ is the set of vertices and $\text{E}\subset{\text{V}}\times{\text{V}}$ denotes the set of edges that are present in the SG. The adjacency matrix of $\textbf{G}$ is represented by $\text{A}$, where each element of this matrix is $+1$ ($-1$) when there is a positive (negative) relation between two nodes, and zero otherwise. We also define $\text{A}^\prime$ to be the positive adjacency matrix whose elements are 1s if there is a link between two nodes, and zero otherwise. Following the notation in~\cite{AM12}, we define the elements of the positive ($\text{P}$) and negative ($\text{N}$) matrices as
\begin{equation}
\text{P}_{ij}=\frac{{A}_{ij}+{A}'_{ij}}{2} \,\,\,\,\,\text{and}\,\,\,\,\,
\text{N}_{ij}=\frac{{A}'_{ij}-{A}_{ij}}{2}\,,
\label{eq:adjacency_relations}
\end{equation}
where the $A_{ij}$ ($A^\prime_{ij}$) are the elements of the adjacency (positive adjacency) matrix and $\{i,j\}\in\{\textbf{V}\}$.
The total number of nodes is $n$ and 
the total number of edges is $m$. The number of non-zero entries in $\text{A}$, $\text{P}$, and $\text{N}$ are denoted by $2\times{m}$, $2\times{m_p}$, and $2\times{m_n}$, respectively. We refer to
the positive degree of vertex $i$ as $p_i$, and its corresponding negative degree as $n_i$. The degree of the vertex $i$ is given by $d_i=p_i+n_i$. We denote a non-empty set of vertices by $\mathscr{C}$ and call it a community cluster.

Our algorithm divides a given community cluster into $k$ communities $\mathscr{C}_1, \mathscr{C}_1, \mathscr{C}_2, \cdots, \mathscr{C}_k$ by optimizing the frustration or modularity as the objective function. Here, we assume that each $\mathscr{C}_l$ ($l\in\{1, 2, \dots, k\}$) is a non-empty set of nodes, and that each node belongs to only one cluster (i.e., there is no overlap between clusters).

When solving for two clusters, we label each node with $s_i$~($i\in\{1,\ldots, n\}$), which as a solution takes the value \mbox{$+1$ ($-1$)} if it falls in the first (second) cluster. We denote $\bf{s}$ as the \mbox{$n$-dimensional} configuration vector and define it as
\begin{equation}
\label{eq:configuration_vector}
{\bf{s}}=[s_1, s_2, \ldots, s_n]\,.
\end{equation}

For multi-cluster detection, we use one-hot encoding to label each node $i$ with one of the $k$ clusters. In particular, we denote the label of a node $i$ by ${\bf{s}}_i$ and define it as 
\begin{equation}
\label{eq:hot-encoding}
{{\bf{s}}_i} = [s_{i1}, s_{i2}, \ldots, s_{ik}]\,,
\end{equation}
where $s_{ic}$ ($c\in\{1,2,\cdots , k\}$) is $1$ if node $i$ belongs to the $c$-th cluster, and zero otherwise. Similar to (\ref{eq:configuration_vector}), we define the \mbox{($k\times{n}$)-dimensional} configuration matrix $\textbf{S}$ as
\begin{equation}
\label{eq:configuration_matrix}
{\textbf{S}} = [\textbf{s}_1^{T}, \textbf{s}_2^{T}, \cdots, \textbf{s}_n^{T}]\,.
\end{equation}

\vspace{0.5em}

\section{Structural Balance}
\label{sec:structural_balance}
The notion of structural balance was first introduced in~\cite{Fri46} to analyze the interaction between pairs of users whose relationships were expressed in terms of being either friends or foes. Later, an SG  approach was taken in~\cite{EK10,WF93} in order to model the social structure between users. The concept of structural balance can perhaps most easily be explained for a simple graph comprising three nodes and then generalized to larger graphs. In what follows, we explain why minimizing the frustration or maximizing the modularity can be appropriate measures to take to find communities in SGs.
\begin{figure}
    \centering
    \includegraphics[width=0.48\textwidth]{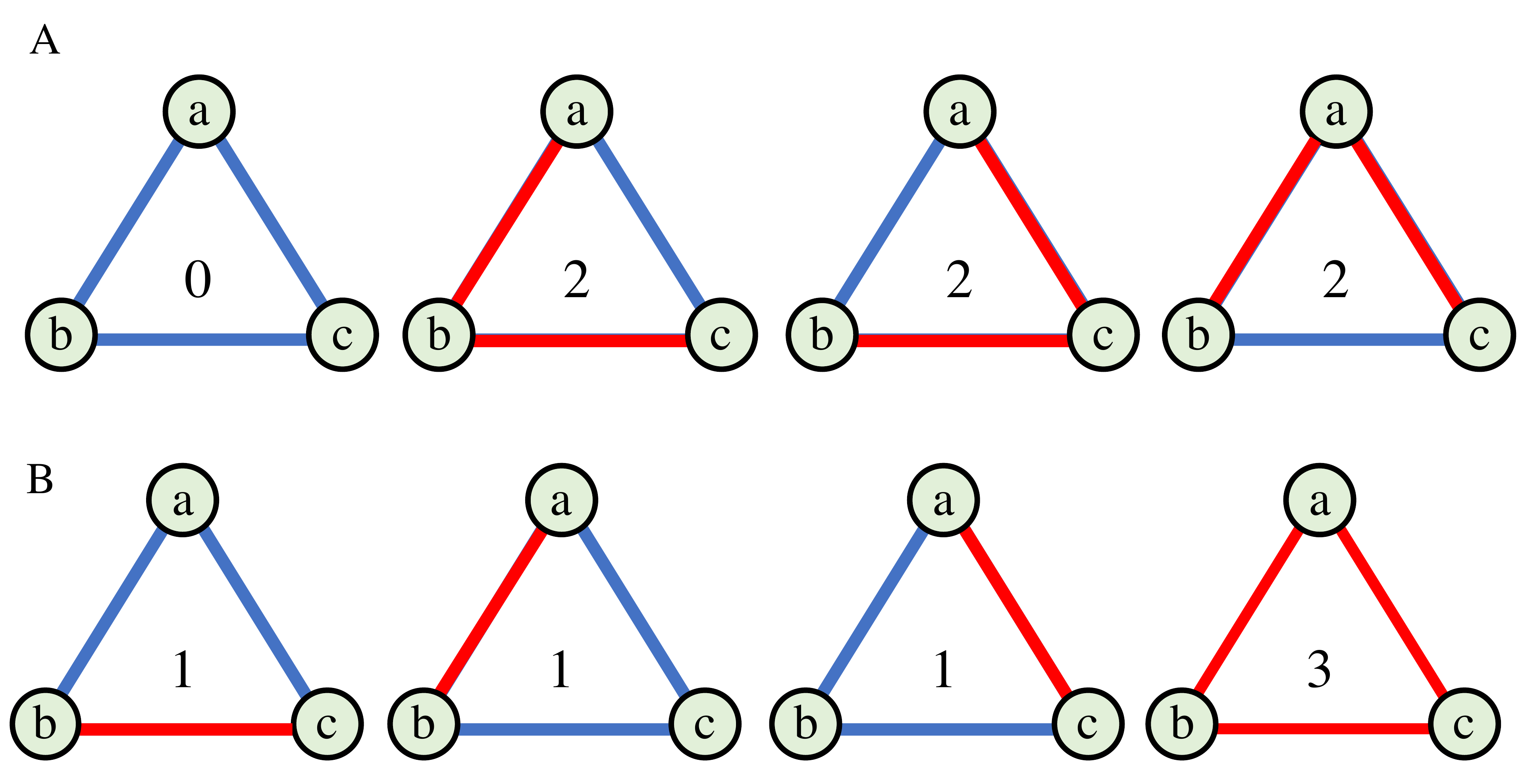}
    \caption{Examples of different configurations of a signed complete graph with three nodes. Each circle represents a node. Nodes are connected via either a negative (a solid line in red) or positive (a solid line in blue) relationship. The numbers inside each graph indicate the total number of negative links in each configuration. The theory of strong social networks defines those configurations with an even (odd) number of negatives edges as structurally stable (unstable).}
    \label{fig:balanced_unbalanced}
\end{figure}

In Fig.~\ref{fig:balanced_unbalanced}, we give an example of an SG with three nodes \{a, b, c\}, where the relation between each pair of nodes can be negative or positive. Assuming the graph is complete (i.e., all nodes are connected to each other by negative or positive links), there exist only four possibilities in general that nodes can be interconnected (i.e., it is a triad). We label each configuration using the total number of negative edges present in that configuration (see Fig.~\ref{fig:balanced_unbalanced}). 

The theory of strong social networks defines those configurations that have an even number of negative links as stable~\cite{harary1953}. Examples of these stable configurations in Fig.~\ref{fig:balanced_unbalanced} are the triad with zero negative links (representing mutual friends) and those with two negative links (representing a pair of friends with a common enemy).
The other two sets of configurations with an odd number of total negative links are unstable, that is, the configurations with one negative tie (representing a pair of enemies with a common friend) and the configuration with three negative ties (representing three mutual enemies). We highlight that there is a generalized structural balance theory proposed in~\cite{Dav67} in which configurations with a total of three negative links are also considered to be stable. Consistent with much other community detection research, we consider here such configurations to be unstable.

We can easily generalize the above discussion to larger graphs by checking that all the fully connected subgraphs of size three in a complete SG are structurally stable. In other words, we call a complete SG \emph{structurally stable} if each 3-clique in SG is structurally stable~\cite{AM12}. In the case of an incomplete SG, we call it \emph{balanced} when it is possible to assign $\pm{1}$ signs to all missing entries in the adjacency matrix, such that the resulting complete network is balanced.

In real-world applications involving SGs, it is often challenging to enforce the notion of structural balance. Instead, the easier concept of clusterizable graphs~\cite{Dav67} is used throughout the literature mostly for solving  community detection problems. In a clusterizable graph, nodes within clusters are connected by only positive links, whereas negative links connect nodes between clusters. Since minimizing the frustration or maximizing the modularity are means for finding clusters in a given network, we use these measures as metrics to find clusters with a high degree of structural balance. In the next section, we give a technical explanation of how frustration or modularity can be used to discover multiple communities in SGs.

\vspace{0.3em}

\section{Methods}
\label{sec:methods}
In this section, we explain our approach to finding multiple communities in social networks by minimizing the frustration or maximizing the modularity within a network. For each of the frustration- or modularity-based approaches, we start with the simpler task of finding two communities in a given SG. We then use the two-community formalism as a building block of our $k$-community detection algorithm.

\subsection{Frustration}
\subsubsection{Two-community detection}
Based on the definition of network frustration, the overall frustration of a given SG is equal to the total number of negatives edges of that network, because all the positive edges lie within a single community (i.e.,~the entire network). Community detection based on the minimization of frustration considers the possibility of any clustering division that leads to a lower frustration contributed by each of the proposed clusters, when compared to the frustration of the full graph considered as a single community. 
Therefore, our goal is to assign a label $s_i$ to each node $i\in\textbf{V}$ such that the resultant assignment lowers (possibly to a minimum value) the frustration within the SG to a value smaller than the frustration of the entire network when considered to be a single community. After such an assignment, each node with a label of $+1$ ($-1$) will belong to the community $\mathscr{C}_1$ ($\mathscr{C}_2$). 

There are two cases that we should consider in order to formulate a measure of frustration. First, any two nodes that are connected by a positive link but are assigned to different communities should increase the frustration. Second, any two nodes which are connected by a negative link but are assigned to the same community should increase the frustration.
We can mathematically express these two cases using the following formula~\cite{AM12} for the frustration $\mathscr{F}$:
\begin{equation}
\label{eq:twocommunity}
\mathscr{F}=\sum_{i,j\in\mathscr{C}}A_{ij} - {\bf{s}}{\bf{A}}{\bf{s}}^T. 
\end{equation}
It is easy to verify that a node pair $(i, j)$ increases 
the value of $\mathscr{F}$ by $1$ if and only if they form a frustrated pair. The solution to the two-community detection problem is then given by the assignment
$\bf{s}^*$ that minimizes $\mathscr{F}$. The minimum frustration will be zero in the case of a fully structurally balanced SG and positive in the case of a partially structurally balanced SG.
We now explain our approach to generalizing this two-community detection algorithm into a \mbox{$k$-community} detection approach.

\subsubsection{Multi-community detection}
As mentioned earlier, we use one-hot encoding to label each node $i$ with one of the $k$ communities. We require that clusters be non-overlapping, that is, each node will be assigned to exactly one cluster. Therefore, we have the following constraint on the label of a given node $i$,
\begin{equation}
\label{eq:norm}
\Vert{\bf{s}}_i\Vert = 1\,,
\end{equation}
where $\Vert\cdot\Vert$ is $l_1$-\text{norm} operator. From  \eqref{eq:hot-encoding} and (\ref{eq:norm}), it follows that if the two nodes $i$, $j$ belong to the same community, we have
\begin{equation}
\label{eq:innerp}
{{\bf{s}}_i}{{\bf{s}}^{T}_j} = 1\,,
\end{equation}
and zero otherwise.

Given (\ref{eq:hot-encoding}--\ref{eq:twocommunity}) and (\ref{eq:innerp}), we can readily generalize the two-community frustration metric (\ref{eq:twocommunity}) into the $k$-community frustration metric $\mathscr{F}^k$: 
\begin{equation}
\mathscr{F}^k=\sum_{i,j\in{\mathscr{C}}}A_{ij} - A_{ij}{{\bf{s}}_i}{{\bf{s}}_j^{T}}\,.
\label{eq:multifrus}
\end{equation}
As we want the detected communities to be non-overlapping, we need to restrict 
the minimization of~\eqref{eq:multifrus} to those configurations ${\bf s}$ that 
are feasible, that is, those that satisfy the constraint that $\Vert {\bf s_i} \Vert = 1$.
Since a QUBO problem is, by definition, unconstrained, we need to transform this 
constrained problem into an unconstrained problem. This is done via a penalty method, using the penalty term
\begin{equation}
\mathscr{P} = M\sum_i(1-\Vert{\bf{s}}_i\Vert)^2\,.
\label{eq:pernalty_t}
\end{equation}
Instead of minimizing the unconstrained objective function \eqref{eq:multifrus}, we then minimize 
the penalized objective function $\mathcal{F}^k_{\mathscr{P}}$:
\begin{equation}
\label{eq:penalized_multi_frustration}
\mathcal{F}^k_{\mathscr{P}} = \mathcal{F}^k + \mathscr{P}\,.
\end{equation}
It is easy to verify that the term inside $\mathscr{P}$ for index $i$ is $0$ if and only if 
$\Vert {\bf s_i}\Vert = 1$, and greater than $1$ otherwise. This means that evaluating 
$\mathscr{P}$ on a feasible solution will yield a value of $0$, and evaluating it 
on an infeasible solution will yield a value of \emph{at least} $M$, which is chosen 
to be a sufficiently large positive number. Details on the selection of $M$ are given in a later section. The effect of this term is to push the objective value of all infeasible solutions higher by at least $M$ while leaving the objective value of a feasible solution unchanged. Thus, if a sufficiently large $M$ is chosen, the optimal solution to 
$\mathcal{F}^k_{\mathscr{P}}$ is guaranteed to be feasible.

To conclude, in~(\ref{eq:penalized_multi_frustration}), we have transformed the \mbox{$k$-community} detection problem into a QUBO problem. An optimal solution ${\textbf{S}}^{*}$, corresponding to the minimum value of $\mathcal{F}^k_{\mathscr{P}}$, will assign each node $i$ to exactly one of $k$ communities. 

\subsection{Modularity}
In this section, we introduce modularity as another methodology for finding multiple communities in a given SG. For unsigned networks, \emph{modularity} is defined as the difference between the number of edges that fall within a community and the number of edges in an equivalent network (i.e., a network with the same number of nodes) when permuted at random~\cite{NG04}. In other words, modularity quantifies a ``surprise" measure which explains the statistically surprising configuration of the edges within the community. Maximizing modularity is then equivalent to having a higher expectation of finding edges within communities compared to doing so by random chance.

In what follows, we first give the formulation for the unsigned and signed networks when the underlying task is to find two communities within a network. We then describe our approach for multi-community detection.

\vspace{0.5em}

\subsubsection{Two-community detection---unsigned graphs}
\label{subsec:twocommunityunsigned}
The notion of modularity has been largely used for detecting communities within unsigned networks (see~\cite{New06}). Let us first consider the two-community detection problem. Without going into details, we use the approach from~\cite{New06} and write the modularity, $\mathcal{M}^\text{u}$, (up to a multiplicative constant) as
\begin{equation}
\label{eq:twoclustermodularity}
\mathcal{M}^\text{u}=\bf{s}\bf{B}^\text{u}\bf{s}^T\,,
\end{equation}
where we have defined the real symmetric matrix $\bf{B}^\text{u}$ as the modularity matrix with the elements
\begin{equation}
\label{eq:modularitymatrix}
B^\text{u}_{ij}=A_{ij}-\frac{d_id_j}{2m}\,,
\end{equation}
where superscript $\text{u}$ in (\ref{eq:twoclustermodularity}) and (\ref{eq:modularitymatrix}) refers to the unsigned graph.  In (\ref{eq:modularitymatrix}), the term $\frac{d_id_j}{2m}$ is the expected number of edges between nodes $i$ and $j$, and all the other symbols have their usual meanings.
Given an optimal configuration $\bf{s}^*$ which maximizes (\ref{eq:twoclustermodularity}), we can assign each node to one of the two communities $\mathscr{C}_1$ and $\mathscr{C}_2$. We next use the approach from~\cite{AM12} to explain how the modularity-based community detection method for unsigned graphs can be expanded into the community detection of SGs.

\vspace{0.5em}

\subsubsection{Two-community detection---signed graphs}
In the case of a signed network, we need to reformulate (\ref{eq:twoclustermodularity}) and (\ref{eq:modularitymatrix})
to include the effect of positive and negative edges. Assuming that our task is to cluster nodes in a given community cluster $\mathscr{C}$ into two clusters $\mathscr{C}_1$ and $\mathscr{C}_2$, we rewrite (\ref{eq:twoclustermodularity}) and (\ref{eq:modularitymatrix})
into the modularity relation, $\mathcal{M}$, for SGs:
\begin{multline}
\label{eq:signedmodulnonmatrix}
\mathcal{M}=\sum_{i,j\in{\mathscr{C}_1}}(P_{ij}-\frac{d_{p_i}{d_{p_j}}}{2m_p})+\sum_{i,j\in{\mathscr{C}_2}}(P_{ij}-\frac{d_{p_i}d_{p_j}}{2m_p}) \\
+\sum_{i\in{\mathscr{C}_1}, j\in{\mathscr{C}_2}}(N_{ij}-\frac{d_{n_i}d_{n_j}}{2m_n})+\sum_{i\in{\mathscr{C}_2}, j\in{\mathscr{C}_1}}(N_{ij}-\frac{d_{n_i}d_{n_j}}{2m_n}).
\end{multline}
Focusing on the right-hand side of (\ref{eq:signedmodulnonmatrix}), we can merge the first two summation terms into a sum over all nodes by multiplying each of the summation terms by 
\begin{equation}
\label{eq:trans1}
\frac{1}{2}(1 + s_is_j)
\end{equation}
and merge the last two summation terms by multiplying each summation term by
\begin{equation}
\label{eq:trans2}
\frac{1}{2}(1 - s_is_j)\,.
\end{equation}
Then, using~\eqref{eq:trans1}--\eqref{eq:trans2}, we can rewrite~\eqref{eq:signedmodulnonmatrix} (up to some constant terms) as:
\begin{equation}
\label{eq:new_signedmodulnonmatrix}
\mathcal{M} = \sum_{i,j\in \mathscr{C}}\left(P_{i,j} - N_{i,j} + \frac{d_{n_i}d_{n_j}}{2m_n} -\frac{d_{p_i}{d_{p_j}}}{2m_p}\right)s_is_j\,.
\end{equation}
Finally, it is straightforward to show that (\ref{eq:new_signedmodulnonmatrix})
can be written in matrix form as
\begin{equation}
\label{eq:signedmodulmatrix}
\mathcal{M}={\bf{s}}{\bf{B}}\bf{s}^T\,,
\end{equation}
where ${\bf{B}}$ is called the ``signed modularity matrix" and, given any two nodes $\{i,j\}\in\bf{V}$, $B_{ij}$ is given by
\begin{equation}
\label{eq:signedfrustrationmatrix}
B_{ij}=A_{ij}+\frac{d_{n_i}d_{n_j}}{2m_n}-\frac{d_{p_i}d_{p_j}}{2m_p}\,.
\end{equation}
All symbols in (\ref{eq:signedmodulnonmatrix}) and (\ref{eq:signedmodulmatrix}--\ref{eq:signedfrustrationmatrix}) have their usual meanings. From \eqref{eq:new_signedmodulnonmatrix} to \eqref{eq:signedmodulmatrix}, we used the relations in \eqref{eq:adjacency_relations} such that $A_{i,j} = P_{i,j} - N_{i,j}$. An optimal configuration $\bf{s}^*$ which maximizes (\ref{eq:signedmodulmatrix}) will assign each node to one of the two communities $\mathscr{C}_1$ and $\mathscr{C}_2$. 

\vspace{0.5em}

\subsubsection{$k$-community detection}
We now generalize the idea of two-community detection to multi-community detection. We explain the core idea for SGs, but note that the method can be easily generalized to the case of unsigned networks.

Our first step in formulating the $k$-community detection algorithm is to generalize
(\ref{eq:signedmodulnonmatrix}) for $k$ communities as follows:
\begin{multline}
\label{eq:kmodularity_1st_step}
    \mathcal{M}^k = \sum_{c=1}^k 
      \sum_{i, j \in \mathscr{C}_c}
      \left(P_{ij} - \frac{d_{p_i} d_{p_j}}{2m_p}\right)
      \\ +
      \sum_{c_1 \not= c_2} 
      \sum_{\substack{i \in \mathscr{C}_{c_1}, \\
            j \in \mathscr{C}_{c_2}}}
      \left( N_{ij} - \frac{d_{n_i} d_{n_j}}{2m_n} \right).
\end{multline}

Following the same approach we took to derive (\ref{eq:signedmodulmatrix}), we can combine the first $k$ summation terms on the right-hand side of (\ref{eq:kmodularity_1st_step}) into a sum over all nodes by multiplying each of the summations by
\begin{equation}
\label{eq:kmodularity_innerproduct_positive}
\textbf{s}^T\textbf{s}
\end{equation}
and merge the last $k(k-1)$ summation terms by multiplying each summation by
\begin{equation}
\label{eq:kmodularity_innerproduct_negative}
1-\textbf{s}^T\textbf{s}\,.
\end{equation}
Using (\ref{eq:kmodularity_innerproduct_positive})--(\ref{eq:kmodularity_innerproduct_negative}) and constraint \eqref{eq:pernalty_t} (for the case of maximization we consider the negative value of $\mathscr{P}$) and the same one-hot encoding approach (\ref{eq:hot-encoding}), we can generalize (\ref{eq:signedmodulmatrix}) to a $k$-community formulation and write the penalized modularity $\mathcal{M}^k_{\mathscr{P}}$ as 
\begin{equation}
\label{eq:modularitymulticommunity}
\mathcal{M}^k_{\mathscr{P}}=\sum_{ij\in\mathscr{C}}B_{{ij}}{\bf{s}}_i{\bf{s}}_j^T- M\sum_i(1-\Vert{\bf{s}}_i\Vert)^2\,,
\end{equation}
where $B_{ij}$ has been defined in (\ref{eq:signedfrustrationmatrix}) and all the other symbols have their usual meanings.
In (\ref{eq:modularitymulticommunity}), we have transformed the \mbox{$k$-community} detection problem into a QUBO problem. An optimal solution, ${\textbf{S}}$, corresponding to the maximum value of $\mathcal{M}^k$ will assign each node $i$ to one of the $k$ communities.

\subsection{Choosing the Hyper-Parameters}
\label{sub:hyper_parameter}
Our $k$-community detection algorithm has two hyperparameters, namely, the maximum number of communities that a user is searching for in the network ($k$) and the penalty coefficient ($M$). In the next two sections, we discuss the role of these two parameters, their features, and suggest a method for predetermining the appropriate values for these parameters.

\vspace{0.5em}

\subsubsection{Penalty coefficient}
In both the frustration and modularity approaches, we penalize configurations that try to simultaneously assign a node to multiple clusters (i.e.,~those configurations which violate the non-overlapping among clusters condition). We do this by adding to (subtracting from) the corresponding objective function of frustration (modularity) a term with a relatively large and positive coefficient $M$. The magnitude of the penalty coefficient depends on the other terms of the objective function.

Since, in our benchmarking examples, we deal with mostly small datasets, we limit our experimental setting to a fixed value for $M$ (see section~\ref{subsubsec:optimizer_setting}). We expect that for larger datasets a fixed value for $M$ will not result in satisfactory performance. Therefore, it becomes necessary to develop a procedure for choosing the appropriate penalty coefficient for each node, based on the structure of the SG, and we do so in what follows. Note that while we derive the method for the case of frustration, it can be applied to the modularity formulation in a similar way.

Thus far in our formulation~\eqref{eq:penalized_multi_frustration}, we have considered the penalty coefficient to have the same value $M$ for each node. Given the topology of the network, different nodes will contribute differently to the objective function. Hence, it is reasonable to have different penalty values for each node. Let us first define the penalty vector $\textbf{M}$ as
\begin{equation}
    \textbf{M} = [M_1, M_2,\cdots, M_n]\,.
\end{equation}
We can then rewrite the second term on the right-hand side of~(\ref{eq:penalized_multi_frustration}) for the case that each penalty coefficient is different for each individual node. Let us call the penalty term $\mathscr{P}$ and write it as
\begin{equation}
\label{eq:penalty_term}
\mathscr{P} = \sum_iM_i(1-\Vert{{\bf{s}}}_i\Vert)^2\,,
\end{equation}
where $i$ iterates over all the nodes in the network and each term has a distinct penalty coefficient term $M_i$. Our goal here is to develop an approach to determining the value of each individual penalty term according to the topology of the SG.

To start, let us consider a feasible solution $\bf{s}_j$ which satisfies the $j$-th constraint in (\ref{eq:penalty_term}) (i.e., the penalty term evaluates to zero). When the constraint is not satisfied by node $j$, the penalty term will contribute a positive value to the total objective function. Changing the assignment $\bf{s}_j$ to 
one that violates the constraint for node $j$, there will be a drop in the first term of the right-hand side of (\ref{eq:penalized_multi_frustration}) (we call this term the unconstrained frustration and denote it by $\mathcal{F}^k$). To find $M_j$ for node $j$, we need to know by how much $\mathcal{F}^k$ will be lowered if we violate the $j$-th constraint imposed solely on node $j$. By setting $M_j$ larger than this amount, we ensure that violating the constraint at 
node $j$ will increase (rather than decrease) the overall 
objective function's value, thus ensuring that the optimal 
solution will satisfy the constraint.

To find an upper bound on the amount the objective 
function can change by violating a constraint, let us consider a node $j$. There are $k$ variables encoding the
communities of node $j$ (see (\ref{eq:hot-encoding})). We check all the terms of $\mathcal{F}^k$ that involve node $j$ and assume that, by violating the constraint, they will lower the $\mathcal{F}^k$ by the maximum possible value of one. Since there are $k \times d_j$ many such drops, we can choose the penalty term for the $j$-th node to be
\begin{equation}
\label{eq:penalty_formula}
M_j := 2d_jk\,,
\end{equation}
where the factor $2$ accounts for the terms resulting from the symmetry of the adjacency matrix of the SG. Using (\ref{eq:penalty_formula}), we can guarantee that the 
optimal solution to the overall objective function 
$\mathcal{F}^k_{\mathscr{P}}$ will
satisfy all constraints.

\vspace{0.5em}

\subsubsection{The number of communities, $k$}
One of the features of our multi-community detection algorithm is that, given an upper bound by the user on the number of communities for which to search, the algorithm will assign each node to one of $k^\prime\leq{k}$ communities, where $k^\prime$ is the optimal number of communities (provided that an optimal solution has been found by the QUBO solver for the corresponding \mbox{$k$-community} detection problem). While our numerical studies demonstrate this feature, we also provide the following argument to support our claim about this feature.

Let us assume that we are given an SG (e.g.,~$\textbf{G}^\prime$) for which we know that $c$ communities exist. In other words, when we assign nodes to $c$ non-overlapping communities, we get the optimal value for the overall frustration or modularity within $\textbf{G}^\prime$. Therefore, we call $c$ the optimal number of communities in $\textbf{G}^\prime$. Now let us encode any given node $i$ in  $\textbf{G}^\prime$ into a \mbox{$k$-dimensional} ($k>c$) $\bf{s}_i$ vector. This means that either (\ref{eq:penalized_multi_frustration}) or (\ref{eq:modularitymulticommunity}) will be formulated to find $k$ communities within~$\textbf{G}^\prime$. If optimizing the frustration or modularity yields more than $c$ non-empty communities, then we have divided at least one of the $c$ communities into sub-communities, increasing the total frustration or decreasing the modularity within the underlying SG. Likewise, when the algorithm returns a number of communities less then $c$, then at least two of the $c$ communities have merged, again increasing (decreasing) the frustration (modularity) within the network. In both cases, neither the solution with more than, nor the solutions with fewer than, $c$ communities is not actually optimal.
Thus, even if the encoding size (which is equivalent to  $k$) provided by the user is larger than the optimal number of  communities, our algorithm returns the optimal number of communities. Note, however, that this argument is only valid if the underlying optimization algorithm returns the optimal solution to the QUBO problem. 

\section{Approach}
\label{sec:approach}
This section discusses our approach to benchmarking our \mbox{$k$-community} detection algorithm. We give a detailed explanation of the datasets, the evaluation criteria for the performance of our algorithm on different datasets, the choice of the QUBO solvers and the optimization approach, and our experimental settings for optimization.

\vspace{0.5em}

\subsubsection{Benchmarking Datasets}
We apply our multi-community detection algorithm on two synthesized datasets and one real-world example dataset~\cite{Rea54}.
We have limited the size of the benchmarking datasets to accommodate the size of the quantum annealer.

\begin{figure}
\includegraphics[width=0.48\textwidth]{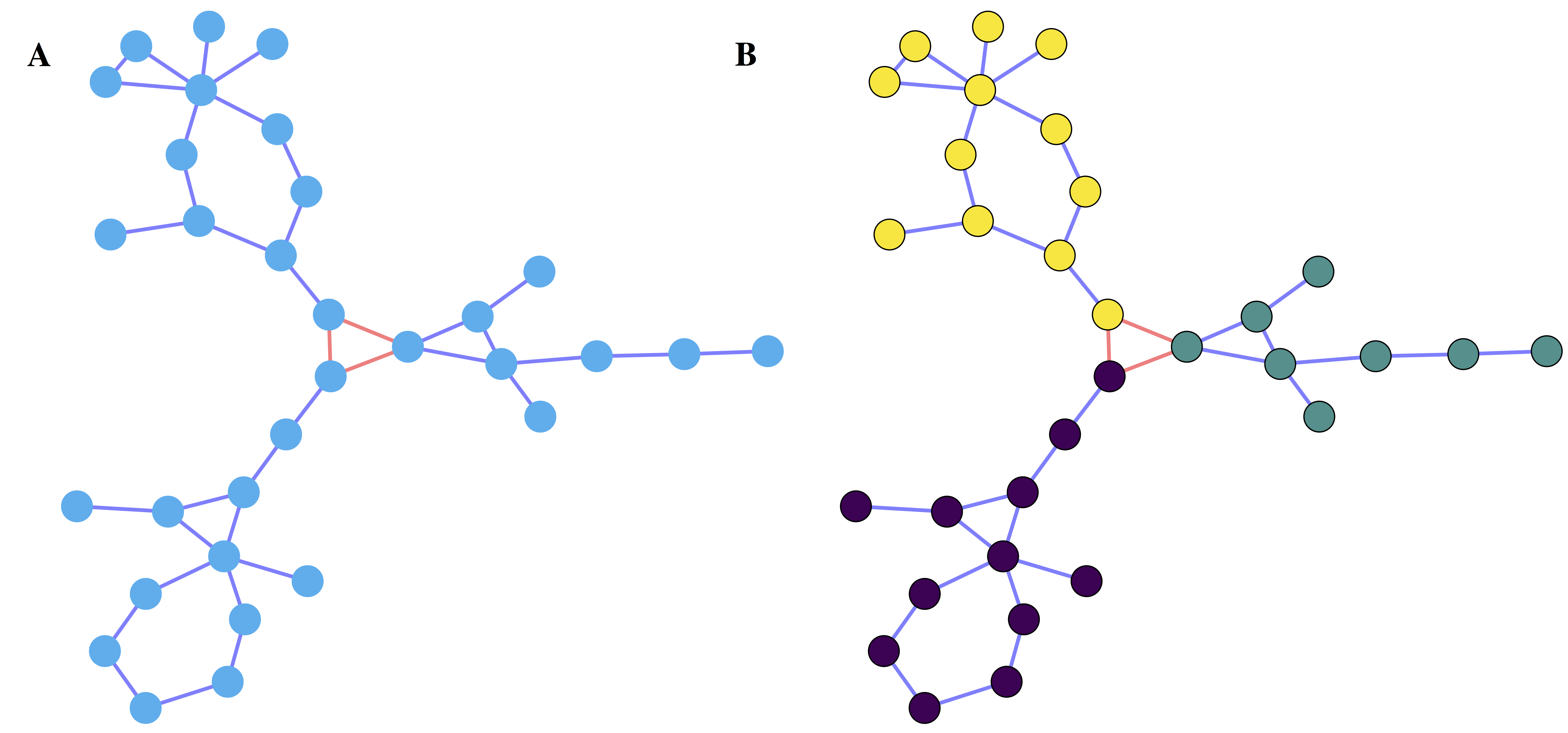}
\vspace{0.8em}
\caption{A randomly generated signed graph ($\mathcal{D}_1$) comprising 32 nodes trivially divided into three clusters of 8, 12, and 12 nodes. Solid circles represent the users that are connected by purple (orange) solid lines, denoting a friendly (antagonistic) relation between them.  A) represents users as being part of one community (all users are represented in blue) which undergoes a graph clustering procedure such that B) each user is assigned to one of three corresponding communities (each community is represented using  green, black, or yellow).}
\label{fig:32nodes}
\end{figure}

\begin{figure}
\includegraphics[width=0.48\textwidth]{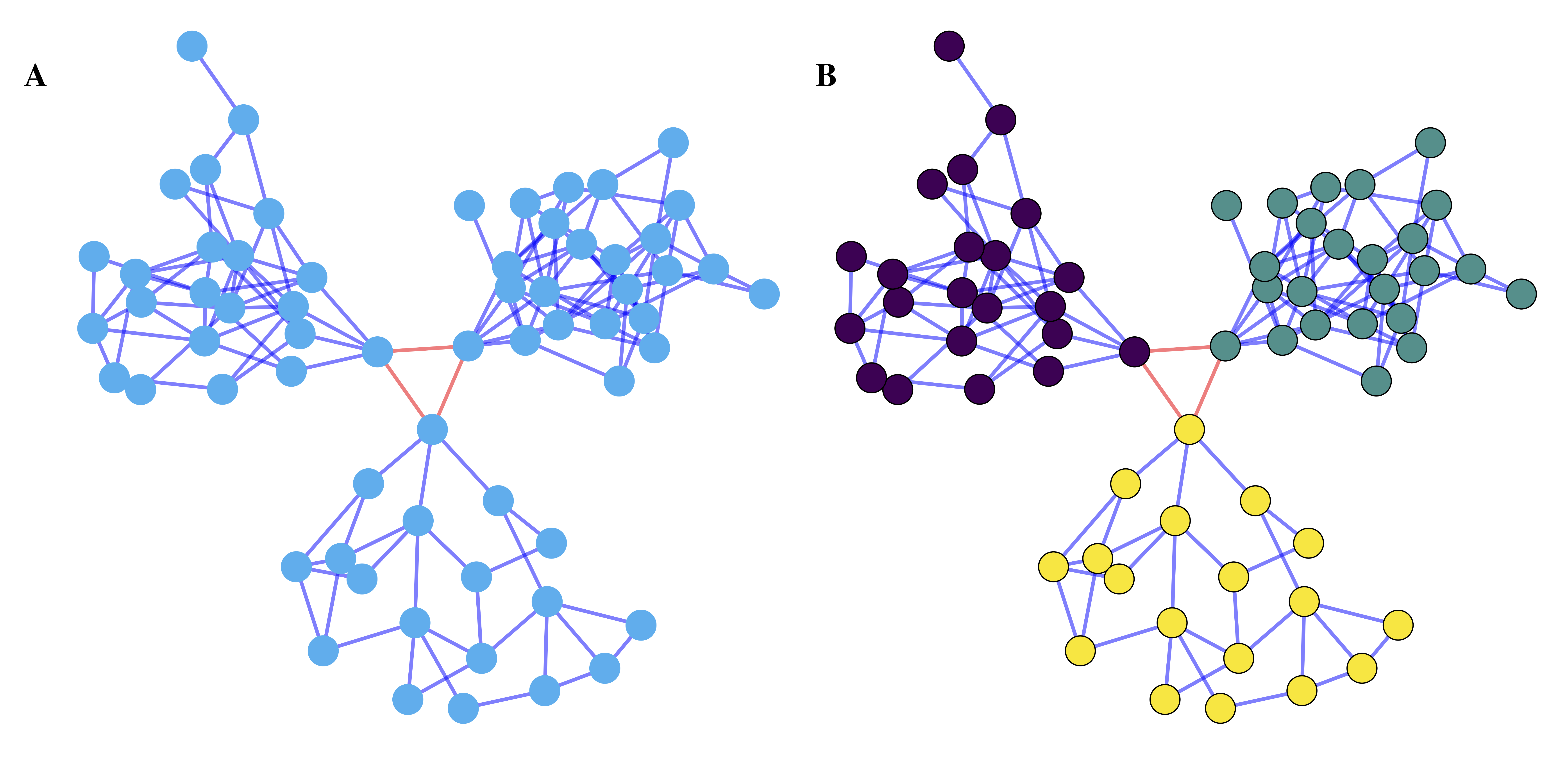}
\caption{A randomly generated signed graph ($\mathcal{D}_2$) comprising 64 nodes trivially divided into three clusters of 18, 22, and 24 nodes. Solid circles represent the users that are connected by purple (orange) solid lines, denoting a friendly (antagonistic) relation between them. A) represents users as being part of one community (all users are represented in blue) which undergoes a graph clustering procedure such that B) each user is assigned to one of three corresponding communities (each community is represented using green, black, or yellow).}
\label{fig:64nodes}
\end{figure}

\begin{figure}
\includegraphics[width=0.5\textwidth]{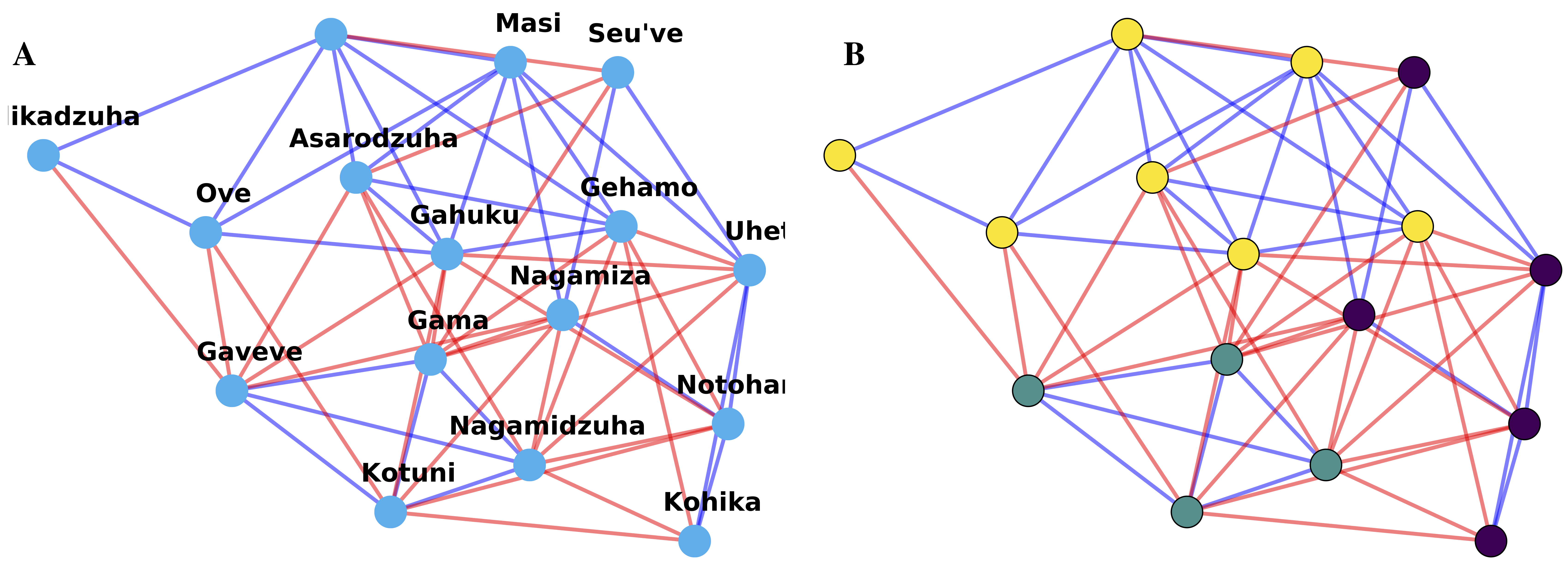}
\caption{An example community detection problem for a real-world example dataset~\cite{Rea54}. The relation between tribal groups (represented using circles) of the Eastern Central Highlands of New Guinea is shown using solid purple (orange) links denoting a friendly (antagonistic) relation between groups. A) represents tribal groups as one community (all tribes are represented in blue) which undergoes a community detection procedure such that B) each tribe is assigned to one of three corresponding communities (each community is represented using green, black, or yellow).}
\label{fig:real_data}
\end{figure}

For the case of synthesized datasets, we consider two SGs of 32 and 64 nodes, where each graph can be trivially clustered into three communities such that the total frustration of the graph is zero. The first synthesized dataset has three clusters of sizes 8, 12, and 12 (Fig.~\ref{fig:32nodes}A). The second graph consists of 64 nodes where three clusters of sizes 18, 22, and 24 form the entire graph (Fig.~\ref{fig:64nodes}A). We employ the following procedure to synthesize the datasets. Given a specified sparsity ($0.2$), we generate three disjoint random clusters where all the nodes within one cluster are positively linked with probability $0.2$. We then choose one node from each community and connect these nodes to each other with negative links, such that we end up with a synthesized dataset with a trivial clustering.
For the case of the real-world example dataset, we consider a dataset~\cite{Rea54} that describes the relation between sixteen tribal groups (represented using solid circles) of the Eastern Central Highlands of New Guinea. Based on previous studies, we know the ground truth (i.e., the optimal number of communities) for this dataset to be three (see Fig.~\ref{fig:real_data}A).

To more easily refer to each dataset in this work, we call the two synthesized datasets with 32 nodes and 64 nodes $\mathcal{D}_1$ and $\mathcal{D}_2$, respectively. We refer to the real-world example dataset with 16 nodes as $\mathcal{D}_3$.

\vspace{0.5em}

\subsubsection{QUBO Solvers}
In (\ref{eq:multifrus})~and~(\ref{eq:modularitymulticommunity}), we have formulated the multi-community detection problems as QUBO problems. We use two QUBO solvers---PTICM and a quantum  annealer (the D-Wave 2000Q). We have chosen PTICM because of its superior performance over a range of other QUBO solvers~\cite{ZOK15,ZFK16}. We also employ a quantum annealer to demonstrate that our algorithm can immediately benefit from advancements in the field of quantum computing.

\vspace{0.5em}

\subsubsection{Block Coordinate Descent}
\label{subsub:BCD}
Given the large size of our dataset with respect to the size of the quantum annealer, we use the decomposition-based approach BCD on top of each QUBO solver to solve QUBO problems whose size exceeds the current capacity of the quantum annealer.

\begin{figure}
    \centering
    \includegraphics[width=0.48\textwidth]{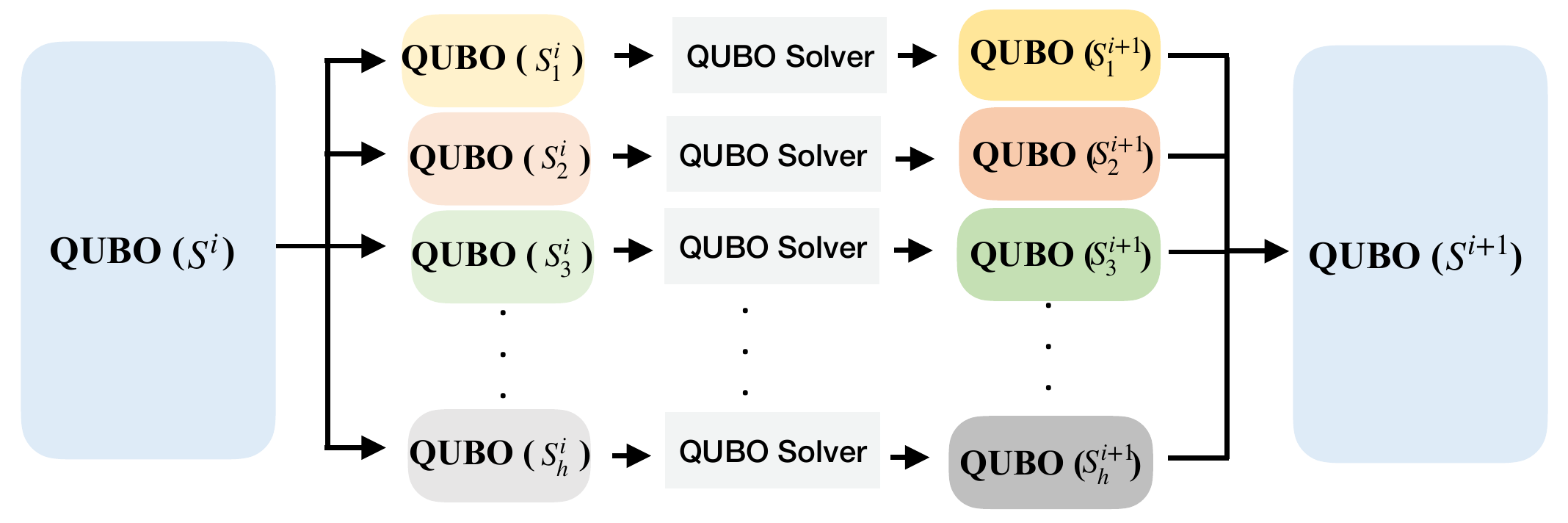}
    \caption{Schematic view of a single iteration of block coordinate descent (BCD). The algorithm first decomposes the entire QUBO problem into $h$ subproblems. Each individual QUBO problem in the reduced space is constructed by fixing the variables that are absent in that subspace. It then employs a QUBO solver algorithm to solve each reduced QUBO problem with $S^i_h$ as the initial configuration (the best feasible solution up to the $i$-th step). At the next step, the algorithm returns the updated solution for each reduced QUBO problem and combines them to build the entire solution of the original QUBO problem such that the updated solution $S^{i+1}$ is the new incumbent solution.}
    \label{fig:BCD}
\end{figure}

BCD works by iteratively solving subproblems while keeping the rest of the variables fixed. We have provided a schematic view of BCD in Fig.~\ref{fig:BCD} that shows different steps of BCD at the $i$-th iteration. To summarize, the algorithm first decomposes the original QUBO problem into a few subproblems by dividing the original variables of the QUBO problem into disjoint subsets. It then uses a QUBO solver to solve each reduced QUBO problem (i.e., subproblem) and updates the solution for each subproblem. Finally, it combines all the subproblems to reconstruct the original QUBO problem with a newly obtained incumbent solution.
Its performance on several QUBO problems is reported in~\cite{RVW+16}. 

\vspace{0.5em}

\subsubsection{Optimizer Settings}
\label{subsubsec:optimizer_setting}
In addition to the hyperparameters of each QUBO solver, the performance of the optimization algorithm for solving the community detection problem is largely dependent on the structure of the graph and the formulation of the QUBO problem. In Section~\ref{sub:hyper_parameter}, we have proposed a formulation for choosing the penalty coefficient when the size of the graph is large. Here we have used a fixed penalty value for each of the community detection problems. Specifically, we have used a penalty value of 10 for the $\mathcal{D}_2$ dataset and 50 for the $\mathcal{D}_1$ and $\mathcal{D}_3$ datasets.

In order to collect statistics on the performance of each QUBO solver on the community detection problems, we run each solver 20 times. Each run includes 200 iterations of BCD (see Fig.~\ref{fig:BCD}). We use a different subspace size for the different problems. For $\mathcal{D}_1$ we choose the value $h=4$ ($h=2$) for the case of frustration (modularity). For $\mathcal{D}_2$, the subspace size is the same ($h=4$) for both frustration and modularity. For the case of the real-world example dataset, we set the subspace size to $h=4$ ($h=5$) for the case of frustration (modularity) (see Section~\ref{subsub:BCD} for more details regarding the BCD and a definition of ``subspace'').

\vspace{0.5em}

\subsubsection{Benchmarking Criteria}
\label{subsec:BC}
To compare the performance of the two  algorithms (i.e., modularity and frustration) we compare the total frustration, of the network after assigning each node to each cluster. We recognize that using frustration might not be a fair metric (especially for large datasets) in comparing the frustration- and modularity-based community detection algorithms. In general, the quality of the solution from each of these algorithms depends largely on the structure of the dataset, and there is no universal method for comparing the performance of different community detection algorithms. However, on small-sized datasets with a trivial structure, we can use frustration as a common means to compare the efficacy of two algorithms. To avoid confusion between the name of the frustration algorithm and our performance metric, we define a criterion of \textit{badness}, $\mathcal{B}$, for each algorithm. The quantity $\mathcal{B}$ is a positive integer which has a lower bound of zero and can take any positive values depending on the structure of a graph. An assignment corresponding to the minimum value for $\mathcal{B}$ is called an optimal assignment, which correctly assigns each node to each community.
We define $\mathcal{P}$ as the success probability of each QUBO solver's finding an optimal assignment over 20 runs.

\vspace{0.5em}

\subsubsection{Multi-Cluster Encoding}
As mentioned earlier, one of the advantages of our proposed multi-community detection algorithm is that it does not require a priori knowledge with respect to the number of communities. For each synthesized dataset, we encode each node into a four-dimensional one-hot encoding vector. This means that, although we know in advance that each of the synthesized datasets has three trivial communities, we still solve the problem as if four communities were present in the given network. A successful community detection algorithm should assign each node to one of the three clusters. In the case of 
the real-world example dataset, we perform the encoding over five clusters and expect the algorithm to assign each node to only three communities.

\vspace{0.5em}

\section{Results}
\label{sec:results}
In this section, we report the results of our experiments on the three datasets considered in this work. We begin with the smaller synthesized dataset, which has 32 nodes ($\mathcal{D}_1$). Fig.~\ref{fig:32nodes}B shows the outcome of an optimal assignment for the case of the $\mathcal{D}_1$ dataset for both the modularity and frustration methods. We also show the optimal assignment of the nodes for the $\mathcal{D}_2$ (Fig.~\ref{fig:64nodes}B) and $\mathcal{D}_3$ datasets (Fig.~\ref{fig:real_data}) for both the modularity and frustration methods.

\begin{figure}
    \centering
    \includegraphics[width=0.23\textwidth]{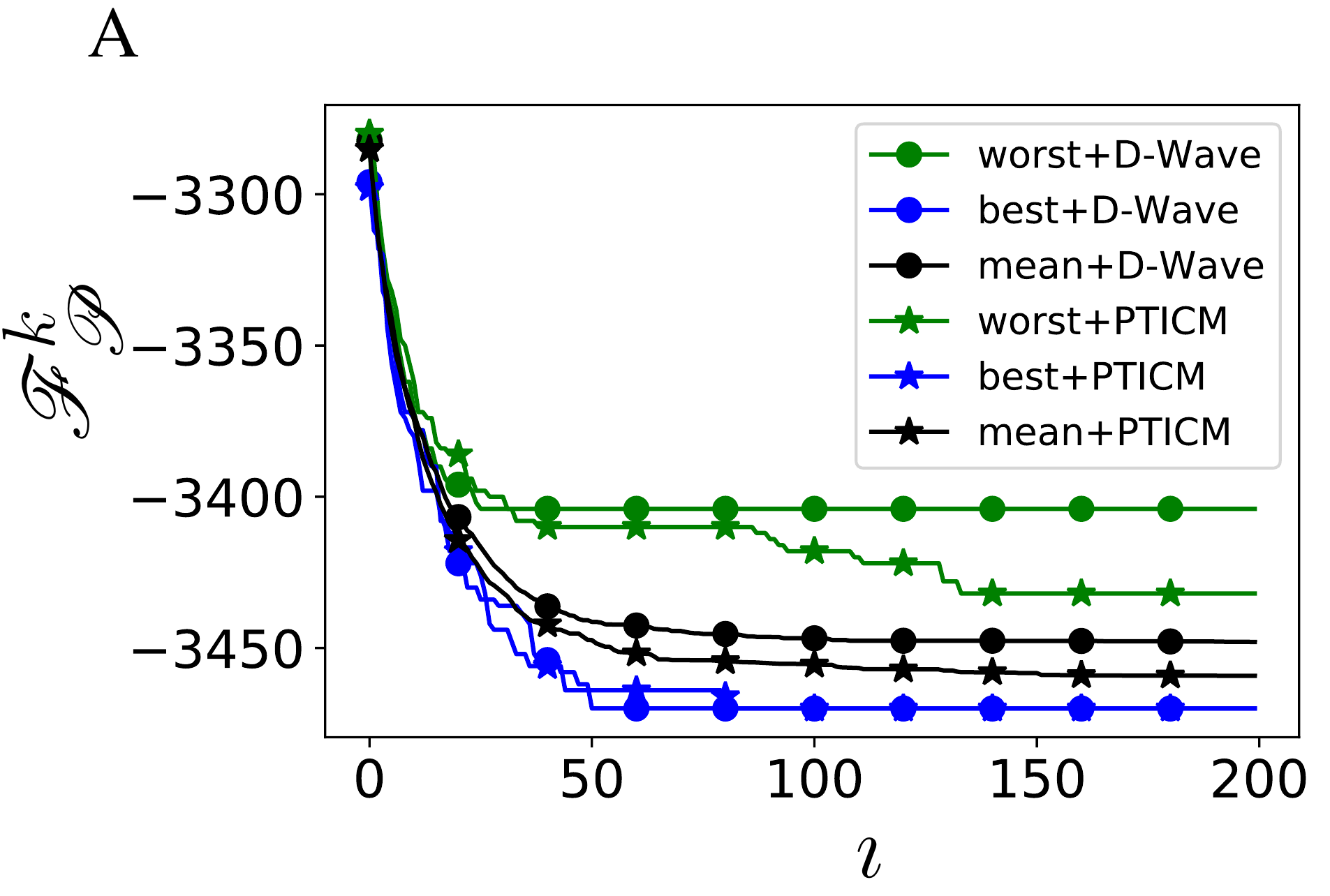}
    \includegraphics[width=0.23\textwidth]{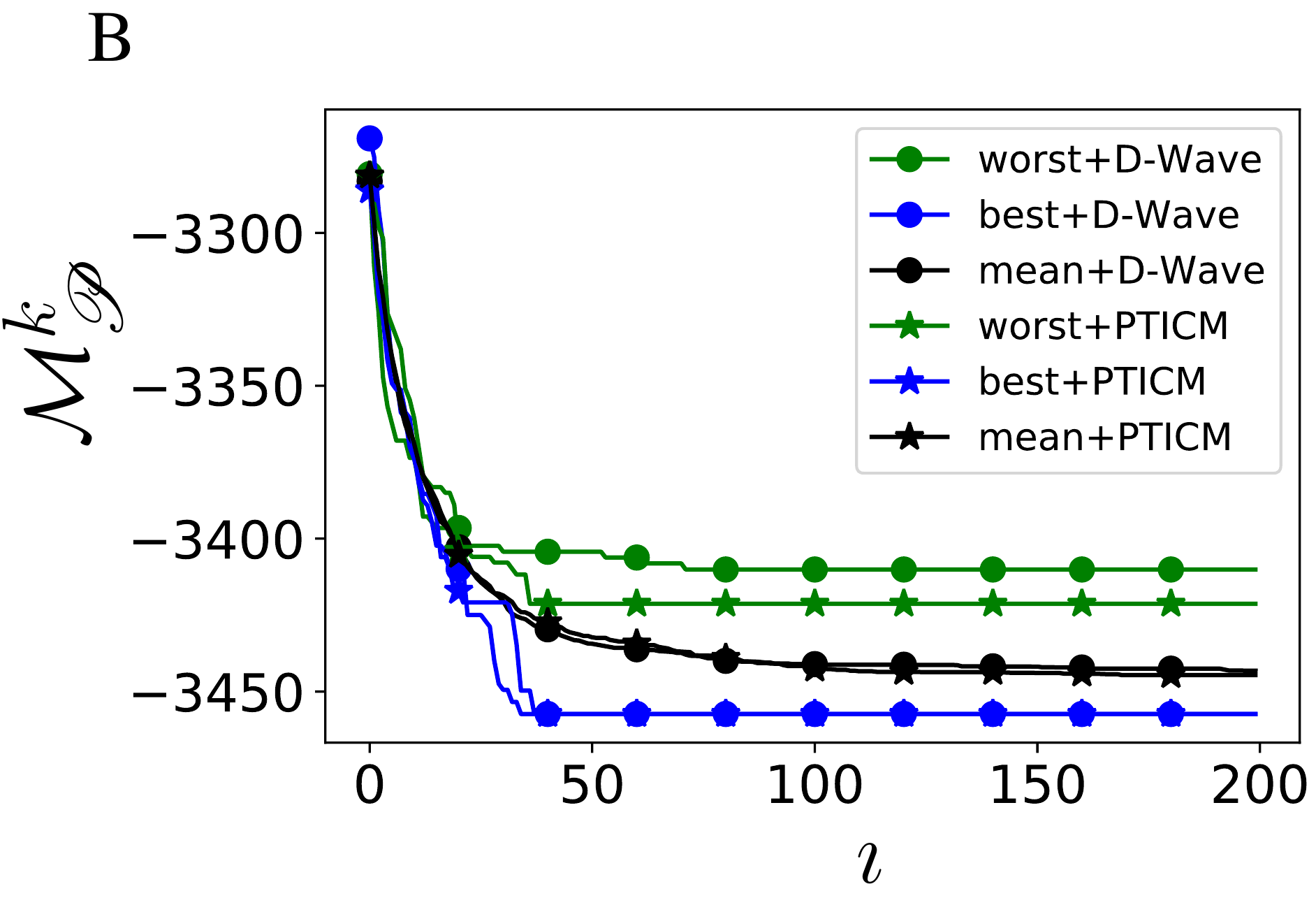}
    \caption{D-Wave 2000Q (solid circle) and PTICM (solid star) results for the 64-node synthesized dataset. We show the worst (green line), best (blue line) and mean (black line) of the A) frustration ($\mathscr{F}^k$) and B) modularity measures vs. the number of iterations ($\imath$) of the BCD algorithm.} 
    \label{fig:energy_convergence}
\end{figure}

To show the convergence of different QUBO solvers on different problems, we have plotted the values of frustration or modularity versus the number of iterations. Fig.~\ref{fig:energy_convergence} shows an example of such a plot for the case of the $\mathcal{D}_2$ dataset.  Fig.~\ref{fig:energy_convergence}A (Fig.~\ref{fig:energy_convergence}B) denotes the results for the case of frustration (modularity). As there are too many such plots to show, we summarize the results of our multi-community detection methods for the other datasets in Tables~\ref{table:frustration_data} and~\ref{table:modularity_data}. 

Specifically, we report the best, the mean, and the worst results for each of the frustration and modularity methods. We report the statistics of these quantities by considering the results at the 200th iteration over 20 runs. We also report the statistics (i.e., $\mathcal{B}_\text{best}$, $\mathcal{B}_\text{mean}$, and $\mathcal{B}_\text{worst}$) of the badness of each method for each solver as well as the success probability $\mathcal{P}$ of each solver to find the optimal assignment of the nodes to multiple clusters for both frustration and modularity.

\begin{table*}[t]
\caption{\vspace{0.7em}
The performance of two QUBO solvers (the D-Wave 2000Q and PTICM) in detecting communities in three signed graphs {~~~~~~~~~~~~} $\mathcal{D}_1$ (a synthesized graph comprising 32 nodes), $\mathcal{D}_2$ (a synthesized graph comprising 64 nodes), and $\mathcal{D}_3$ (a real-world example graph comprising 16 nodes). Here we use frustration as a metric for finding communities. The best, mean, and worst quantities are taken from the last iteration of each of the 20 runs of each QUBO solver. $\mathcal{B}$ refers to the badness of the frustration-based algorithm (see~Section~\ref{subsec:BC}), where a lower value of $\mathcal{B}$ denotes better performance of the frustration method over the modularity (see Table~\ref{table:modularity_data}) method. $\mathcal{P}$ is the success probability of each QUBO solver finding the optimal solution for each \mbox{$\mathcal{D}_i$ ($i\in\{1,2,3\}$)} over the $20$ runs for each QUBO solver.}
\label{table:frustration_data}
\centering
\begin{tabular}{rcccccccc}\toprule
\multicolumn{9}{c}{Frustration ($\mathcal{F}^k_\mathscr{P}$)}                                      \\ \otoprule
\multicolumn{1}{c}{}                    &       & best & mean & worst & $\mathcal{B}_\text{best}$ & $\mathcal{B}_\text{mean}$ & $\mathcal{B}_\text{worst}$ & $\mathcal{P}$  \\ \otoprule
\multicolumn{1}{l}{\multirow{2}{*}{$\mathcal{D}_1$}} & D-Wave Solver& $-388$  & $-382$ & $-378$ & $0.0$ & $2.9$ & $5.0$ & $0.10$ \\ \cmidrule(l){2-9}
\multicolumn{1}{l}{}                    & PTICM & $-388$ & $-386$ & $-382$ & $0.0$ & $0.7$ & $3.0$ & $0.45$ \\ \midrule
\multirow{2}{*}{$\mathcal{D}_2$}                     & D-Wave Solver & $-3470$ & $-3448$ & $-3404$ & $0.0$ & $11.0$ & $33.0$ & $0.05$  \\ \cmidrule(l){2-9} 
                                        & PTICM & $-3470$ & $-3459$ & $-3432$ & $0.0$ & $5.4$ & $19$ & $0.05$  \\ \midrule
\multirow{2}{*}{$\mathcal{D}_3$}                     & D-Wave Solver& $-374$ & $-374$ & $-374$ & $2.0$ & $2.0$ & $0.0$ & $1.0$ \\ \cmidrule(l){2-9} 
                                        & PTICM & $-374$ & $-374$ & $-374$ & $2.0$ & $2.0$ & $0.0$ & $1.0$ \\ \bottomrule
\end{tabular}
\end{table*}

\begin{table*}[t]
\caption{\vspace{0.7em}The performance of two QUBO solvers (the D-Wave 2000Q and PTICM) in detecting communities in three signed graphs {~~~~~~~~~~~~}  $\mathcal{D}_1$ (a synthesized graph comprising 32 nodes), $\mathcal{D}_2$ (a synthesized graph comprising 64 nodes), and $\mathcal{D}_3$ (a real-world example graph comprising 16 nodes). Here we use modularity as a metric for finding communities. The best, mean, and worst quantities are taken from the last iteration of each of the 20 runs of each QUBO solver. $\mathcal{B}$ is called the badness of the modularity-based algorithm, where a lower value of $\mathcal{B}$ denotes better performance of the modularity method over the frustration (see Table~\ref{table:frustration_data}) method. $\mathcal{P}$ is the success probability of each QUBO solver finding the optimal solution for each \mbox{$\mathcal{D}_i$ ($i\in\{1,2,3\}$)} over the $20$ runs for each QUBO solver.\label{table:modularity_data}}
\centering
\begin{tabular}{rcccccccc}\toprule
\multicolumn{9}{c}{Modularity ($\mathcal{M}^k_{\mathscr{P}}$)}                                      \\ \otoprule
\multicolumn{1}{c}{}                    &       & best & mean & worst & $\mathcal{B}_\text{best}$ & $\mathcal{B}_\text{mean}$ & $\mathcal{B}_\text{worst}$ & $\mathcal{P}$  \\ \otoprule
\multicolumn{1}{l}{\multirow{2}{*}{$\mathcal{D}_1$}} & D-Wave Solver & $-384$  & $-383$ & $-381$ & $0.0$ & $0.5$ & $2.0$ & $0.5$ \\ \cmidrule(l){2-9} 
\multicolumn{1}{l}{}                    & PTICM & $-384$ & $-383$ & $-381$ & $0.0$ & $0.5$ & $2.0$ & $0.50$ \\ \midrule
\multirow{2}{*}{$\mathcal{D}_2$}                     & D-Wave Solver & $-3457$ & $-3443$ & $-3410$ & $0.0$ & $7.5$ & $25.0$ & $0.05$  \\ \cmidrule(l){2-9} 
                                        & PTICM & $-3457$ & $-3444$ & $-3421$ & $0.0$ & $6.8$ & $19$ & $0.05$  \\ \midrule
\multirow{2}{*}{$\mathcal{D}_3$}                     & D-Wave Solver & $-852$ & $-852$ & $-852$ & $2.0$ & $2.0$ & $2.0$ & $1.0$ \\ \cmidrule(l){2-9} 
                                        & PTICM & $-852$ & $-852$ & $-852$ & $2.0$ & $2.0$ & $2.0$ & $1.0$ \\ \bottomrule
\end{tabular}
\end{table*}

\vspace{0.5em}

\section{Discussion}
\label{sec:discussion}
In this section, we discuss the performance of our algorithms on the three benchmarking datasets that we have considered in this work. We begin with Fig.~\ref{fig:32nodes}A, which shows the original SG $\mathcal{D}_1$, where all nodes are assigned to one community. It is trivial for any successful community detection algorithm to discover the three individual communities within $\mathcal{D}_1$. We take a different approach from existing divisive-based multi-community detection algorithms to solve this problem. One of the advantages of our proposed method is that the algorithm requires only an upper-bound estimate of the number of communities from the user to search for the optimal number of communities. To demonstrate this, we initially assume that there exist four communities ($k=4$) within the graph. The number of variables in the underlying QUBO algorithm increases linearly with $k$ and is proportional to $n\times{k}$. Therefore, we have to solve a QUBO problem with 132 binary variables. Fig.~\ref{fig:32nodes}B shows that, despite the initial assumption that there are four communities within the graph, both frustration- and modularity-based methods report three communities as the optimal number of communities.

We have followed the same approach for the case of $\mathcal{D}_2$. We run the algorithm with the assumption that there are four communities ($k=4$) within the graph. For this case, we have to solve a QUBO problem with \mbox{$4\times{64} = 256$} variables. Fig.~\ref{fig:64nodes}B shows the results of community detection for both the frustration and modularity methods. Both methods discover three communities despite the initial assumption of there being four communities. For the case of the real-world example dataset (Fig.~\ref{fig:real_data}A), we solve the problem such that we assume five communities within the graph. The corresponding QUBO problem size is $5\times{16} = 80$. Both the modularity and frustration methods correctly assign each node to one of three communities.

To show that our method can benefit immediately from progress within quantum technology, we have considered the \mbox{D-Wave 2000Q} quantum annealer as one of the QUBO solvers. This quantum annealer can solve a community problem for a complete SG of size 64. Since the size of all the QUBO problems that we consider in this work exceeds the current capabilities of available quantum annealers, we employ the BCD algorithm to iteratively solve each subproblem of the original QUBO problem. In general, our community detection algorithm does not rely on the BCD algorithm and can be used without using BCD so long as the underlying computing device can handle the size of the corresponding QUBO problem.

We have summarized the performance of the two QUBO solvers on various benchmarking problems in Tables~\ref{table:frustration_data} and~\ref{table:modularity_data}. Table~\ref{table:frustration_data} reports the results for the case of frustration ($\mathscr{F}^k_\mathscr{P}$) and Table~\ref{table:modularity_data} reports the results for the case of modularity ($\mathcal{M}^k_\mathscr{P}$). We begin with the results pertaining to the frustration method.

The first column of Table~\ref{table:frustration_data} shows that both the D-Wave solver and PTICM are able to find the optimal number of communities within all three datasets. This is also clear from the column that shows the $\mathcal{B}_{\text{best}}$ for different datasets. The badness measure for the first two datasets ($\mathcal{D}_1$ and $\mathcal{D}_2$) are zero, which corresponds to a perfect assignment of each node of these graphs to one of the three communities (see Fig.~\ref{fig:32nodes}B and Fig.~\ref{fig:64nodes}B). For the case of the real-world example dataset $\mathcal{D}_3$, the optimal assignment of the nodes to three individual communities corresponds to a badness of $2.0$, as the nontrivial structure of the network does not allow any assignment with $\mathcal{B}=0$.

The results from the mean, worst, $\mathcal{B}_\text{mean}$, and $\mathcal{B}_\text{worst}$ columns confirm the superior performance of PTICM over the D-Wave solver for the case of frustration-based method. Over 20 runs, PTICM could find the optimal number of communities in 45\% and 15\% of the time for the case of $\mathcal{D}_1$ and $\mathcal{D}_2$, whereas these quantities are 10\% and 5\% for the D-Wave solver. Both algorithms perform the same on $\mathcal{D}_3$, with a success probability of $\mathcal{P}=1.0$.  We would like to stress that our results regarding the superiority of PTICM over the D-Wave solver are not conclusive, as we had  resources insufficient for performing a proper hyper-parameter tuning on the D-Wave quantum annealer.

We now discuss Table~\ref{table:modularity_data}, which includes the modularity ($\mathcal{M}^k_\mathscr{P}$) results. From the first column of the table, we can see that both the D-Wave and PTICM solvers find the optimal number of communities for all three datasets. Looking at $\mathcal{B}_\text{mean}$, we can see that the performance of PTICM and the D-Wave solver is comparable on all three datasets. For each dataset, both solvers find the optimal number of communities with almost the same success probability.

\section{Concluding Remarks and Future Work}
\label{sec:conclusion}
In this work, we have devised a multi-community detection algorithm to find communities within signed graphs. We have tested our approach on three different datasets, including two randomly generated signed graphs and one real-world dataset. Our approach has two main advantages over existing algorithms. First, when searching for  multiple communities within a graph, our algorithm preserves the global structure of the network as opposed to divisive community detection algorithms, which only consider the local structure. Second, our method does not require a priori knowledge of the number of communities. Given an upper bound on the number of communities, our algorithm will find the optimal number of communities, which will be equal to or less than the predefined upper bound chosen by the user.

Our main focus in this work has been to provide the formulation of a multi-community algorithm that  uses frustration or modularity as a means to detect communities. The next step would be to test our method on larger datasets using available state-of-the-art classical optimizers such as the Digital Annealers~\cite{ART+18}. 

\section{Acknowledgement}
The authors thank Marko Bucyk for reviewing and editing the manuscript, and Moslem Noori and Sourav Mukherjee for helpful and constructive discussions. This work is fully funded by 1QBit.

\clearpage
\bibliographystyle{naturemag}

\end{document}